\documentstyle[aps,twocolumn,epsfig]{revtex}
\begin{document}
\draft
\date{\today}
\title{Non-extensive thermodynamics and
stationary processes of localization}
\author{Leone Fronzoni$^{1,2}$, Paolo Grigolini$^{3,4,5}$,
Simone Montangero$^{1}$}

\address{$^{1}$Dipartimento di Fisica dell'Universit\`{a} di Pisa and INFM,
Via Buonarroti 2, 56127 Pisa, Italy }
\address{$^{2}$Centro interdisciplinare per lo studio dei Sistemi Complessi
Via Bonanno 25B, Pisa, Italy }
\address{$^{3}$Dipartimento di Fisica dell'Universit\`{a} di Pisa,
Piazza
Torricelli 2, 56127 Pisa, Italy }
\address{$^{4}$Center for Nonlinear Science, University of North
Texas, P.O. Box 305370, Denton, Texas 76203 }
\address{$^{5}$Istituto di Biofisica del CNR, Via San Lorenzo 26, 
56127 Pisa, Italy }
\maketitle

\begin{abstract}
We focus our attention on dynamical
processes characterized by
an entropic index $Q<1$. According to the probabilistic
arguments of Tsallis and Bukman
[ C. Tsallis,D.J. Bukman, Phys.Rev.E {\bf
54}, R2197 (1996)] these processes
are subdiffusional in nature. The
non-extensive generalization
of the Kolmogorov-Sinai
entropy yielding the same entropic index
implies the stationary condition.
We note, on the other hand, that enforcing the stationary property
on subdiffusion  has
the effect of producing a localization
process occurring within
a finite time scale. We thus conclude
that the stationary dynamic processes
with $Q<1$ must
undergo a localization process occurring at
a finite time.
We check the validity of this
conclusion  by means of a
numerical
treatment of the dynamics of the logistic map
at the critical point.
\end{abstract}


\section{introduction}
There is a growing interest\cite{list} on the subject of non-extensive
thermodynamics introduced ten years ago by
 Tsallis \cite{CONSTANTINO88}, by means of the entropy
 \begin{equation}
 S_{q}= \frac{1 - \sum_{i = 1}^{W}p_{i}^{q}}{q-1}.
 \label{entropy}
 \end{equation}
 Note that this entropy is characterized by the index $q$ whose
 departure from the
 conventional value $q = 1$ signals the thermodynamic effects of
 either long-range correlations in fractal dynamics
 \cite{heuristic1,heuristic2,heuristic3} or the non-local
 character of quantum mechanics\cite{luigi,rajagopal}.The growing interest for
 Tsallis' non-extensive entropy
 is testified by the exponentially
 growing list of publications on this hot issue\cite{list}.
 Of remarkable interest for the subject of fractal dynamics is the
 discovery recently made
 in Refs.\cite{heuristic1,heuristic2,heuristic3}
 that
 the entropic index $q$ also determines the specific analytical
 form, more general than the exponential form, adopted by two
 trajectories moving from infinitelly close initial conditions, to
 depart from one another.

  This important result is based
 on the generalization of the Kolmogorov-Sinai (KS) entropy \cite{K,S}
 and, consequently, of the important theorem of Pesin \cite{pesin}.
In Refs.~\cite{heuristic1,heuristic2,heuristic3} it was shown that the
adoption of
the Tsallis entropy yields a natural generalization of the
Kolmogorov-Sinai (KS)
entropy \cite{ruelle}. For the sake
of simplicity, henceforth
we shall be referring ourselves
to this form
of entropy
as Kolmogorov-Sinai-Tsallis (KST) entropy.
The adoption of the KST entropy yields a generalization of the
Pesin theorem \cite{heuristic1,heuristic2,heuristic3} and, with it,  the
important result:
\begin{equation}
\delta(t) \equiv \lim_{\Delta y(0) \rightarrow 0} \frac{\Delta
y(t)}{\Delta y(0)}
= [1 + \lambda_{q}t(1-q)]^{1/(1- q)}.
\label{delta(t)}
\end{equation}
The symbol $\Delta y(t) $ denotes the distance at time $t$ between
two trajectories departing from two close initial conditions.
Note that this
prescription for $q>1$ implies a divergence at finite time, and,
consequently, a
faster than exponential departure of two
initially very close trajectories from one to the other.
The case $q<1$ yields a power law
dependence on time, which can be interpreted as being slower than the
exponential departure. It is well known that the case $q=1$ results
in ordinary Brownian diffusion. In Ref.~\cite{condmatt} it was shown
that the case $q>1$ leads to superdiffusion. Consequently, the case
$q<1$, with a departure of the trajectories slower than exponential,
ought to result,
in general, in subdiffusion. It has to be pointed out that
this dynamical
argument fits the theoretical prediction of Ref.~\cite{tsallisbukman96}. The
authors of this interesting paper assign an entropic index $q$
to a non-linear Fokker-Planck equation, and prove
that the condition $q<1$ generates a subdiffusion process.
We are convinced that there is an intimate relation between
dynamic and probabilistic approach\cite{polish,bianucci}; as a
consequence of this conviction we are led to believe that also a dynamic
process characterized by $q<1$ must produce a form of
subdiffusion.

 As well known\cite{ruelle,pincus}, the concept of KS entropy
 implies the assumption of an invariant distribution. This condition, which
  must
 be extended to the KST entropy
 is ,  in turn, closely related to the stationary assumption recently
 adopted\cite{mannella} to establish a dynamic derivation of
 the diffusion processes. We want to prove that this assumption is
 the key physical ingredient necessary to make the dynamic treatment
 of
 anomalous diffusion compatible with a treatment based on
 non-extensive thermodynamics. This has also the effect of making
 subdiffusion as well as superdiffusion\cite{condmatt} compatible with
 the extended thermodynamics of Tsallis.

 On the other hand, in Section II we will show that
 the only form of subdiffusion compatible with the
 stationary condition is a localization process taking
 place with a finite time scale. For all these reasons
 we are led to make the conjecture that a generator of diffusion
 characterized by the entropic index $q < 1$ must result
 in a localization process, whose time scale is finite,
 and it is expected to tend to infinity with $q \rightarrow 1$.

 We shall not produce a general proof of this conjecture.
 We shall limit ourselves to
 a numerical investigation
 of a dynamical generator
 with $q<1$ and we shall show, using in fact a numerical approach,
 that it results in a localization process.
  The dynamic generator of diffusion studied in this paper
 is the logistic map.
 The rationale for this choice is that the logistic map is one of the
 most used prototypes to illustrate the transition to chaos
 \cite{chinesebook}. On top of that, there are already
 some papers\cite{heuristic1,heuristic2,heuristic3}
 devoted to studying the sensitivity to initial condition in the case
 of the logistic map, and claiming that the proper entropic
 index in this case is $q <1$. Thus, to prove our conjecture
 that the stationary dynamics with $q<1$ yield localization,
 we are only left with the problem of proving, with numerical
 calculations, that the
 diffusion process generated by the logistic map
 is in fact characterized by localization.

 The paper is organized as follows. In Section II we review the
 dynamic approach to diffusion, a picture based on very simple
 mathematical arguments. We show that the condition that anomalous
 diffusion is  slower than normal diffusion, and at the same time
 compatible with the stationary assumption, yields localization
 occurring within a finite time scale. Section III is devoted to
 discussing the time evolution of the KST entropy of the logistic map
 at the onset of chaos, and to the detection of the entropic index
 $q<1$.
 In Section IV we check the theoretical conjecture that
 $q <1$ yields localization with a finite time scale.
 We study also the localization breakdown caused
 by the control parameter approaching the condition of full chaos.
 Finally, Section V is devoted to concluding remarks.

\section{a dynamic approach to diffusion}
The simplest dynamical approach to diffusion is established by the
equation of motion\begin{equation}
\frac{dx}{dt} = \xi(t).
\label{mainequation}
\end{equation}
The variable $x$ is the diffusing variable and at a given time $t>0$
we have\begin{equation}
x(t) = \int_{0}^{t}\rm d$t$^{\prime}\xi(t^{\prime}) + x(0).
\label{x(t)}
\end{equation}
Following \cite{mannella} we make first of all the assumption that the
time evolution of $\xi(t)$ corresponds to a stationary statistical
condition, which, in the specific case of the logistic map studied in
this letter, implies the existence of an invariant
distribution, whose structure will be studied numerically in Section
IV. In this specific condition it is straightforward to
prove\cite{mannella} that\begin{equation}
\frac{d\langle x^{2}(t)\rangle}{dt}
 = 2 D(t),
\label{2D(t)}
\end{equation}
where $D(t)$ can be interpreted as being a time dependent diffusion
coefficient defined by\begin{equation}
D(t) \equiv
\langle \xi(t)^{2}\rangle_{eq}\int_{0}^{t} \Phi_{\xi}(t^{ \prime})
dt^{ \prime}
\label{defineD(t)}
\end{equation}
and the symbol $\Phi_{\xi}(t)$ denotes the autocorrelation
function of the fluctuation $\xi(t)$ defined by\begin{equation}
\Phi_{\xi}(t) \equiv
\frac{\langle \xi(0)\xi(t)\rangle_{eq}}
{\langle \xi^{2} \rangle_{eq}}.
\label{correlationfunction}
\end{equation}
Note that the stationary property implied by the one-time
correlation function of
Eq.(\ref{correlationfunction}) is generated by averaging over an equilibrium
distribution, or invariant measure, of initial condition.
Ordinary statistical mechanics is ensured by the condition that the
correlation time $\tau$ defined by\begin{equation}
\tau \equiv
\int_{0}^{t} \Phi_{\xi}(t^{ \prime})
dt^{ \prime}
\label{tau}
\end{equation}
exits and is finite, namely, that the important
condition\begin{equation}
0 < \tau < \infty
\label{importantcondition}
\end{equation}
is fulfilled.

The important condition of
Eq.(\ref{importantcondition}) can be violated in different ways.
A first example is given by a physical condition described by
the autocorrelation function $\Phi_{\xi}(t)$
 with the time
asymptotic property
\begin{equation}
\lim_{t\rightarrow\infty}\Phi_{\xi}(t )
= \frac{const}{t^{\beta }}
\label{asymptotics}
\end{equation}
and

\begin{equation}
0 < \beta < 1.
\label{anotherimportantcondition}
\end{equation}
This corresponds to the time dependent diffusion coefficient of
Eq.(\ref{defineD(t)}) becoming infinite in the asymptotic time limit.
Another way of violating this condition, on wich we focus our attention in 
this paper, is when the time $\tau$ of
Eq. (\ref{tau}) vanishes. In this case, in the asymptotic time
limit  the time dependent diffusion
coefficient of Eq.({\ref{defineD(t)}) vanishes,
thereby implying that also the localization
process might take place asymptotically
in time.

The methods of nonequilibrium statistical mechanics
can be extended to deal with
processes violating the important condition of
Eq.(\ref{importantcondition}).
Examples can be found in the recent
literature, see for instance \cite{giorgio}, and in interesting review
papers\cite{geisel,klafter}.  Furthermore, as an effect of
the increasing popularity of the non-extensive entropy proposed
ten year ago by Tsallis\cite{CONSTANTINO88} it is becoming clear that
also the regime of anomalous diffusion can be given a thermodynamical
as well a dynamical significance\cite{condmatt}.

We aim at establishing a unified mechanical and
 thermodynamical picture ranging from normal to anomalous diffusion,
 namely,
 including both super and subdiffusion. For this
 ambitious task to be succesfully completed
 it is necessary to turn our attention to the
 stationary property. First of all, we have to point out that the
 analysis made in Ref.~\cite{condmatt} was limited to the case where
 the fluctuating variable $\xi(t)$ is dichotomous, and the map of
 Geisel and Thomae \cite{geiselthomae84} is used to establish the
 distribution of sojourn times in one or the other of the velocity
 states \cite{allegrini}. This waiting time distribution, called
 $\psi(t)$, has
 the inverse power law structure:
 \begin{equation}
 \lim_{t \rightarrow \infty}
 \psi(t)
= \frac{const}{t^{\mu}}
\label{psi(t)}
\end{equation}
with
 \begin{equation}
 2 < \mu <3 .
\label{restriction}
\end{equation}
The restriction to values of $\mu > 2$ is enforced by the need of
establishing a contact with the stationary theory earlier developed.
In fact, as recently pointed by Zaslavsky\cite{zaslavsky}, in the
specific case when the
waiting time distribution has an inverse power behavior, the asymptotic
time regime of the distribution of Poincar\'{e} recurence time is
dominated by the distribution of times of sojourn at the border between
the chaotic sea and the stability islands. On the other hand the Kac
theorem\cite{KAC} establishes that ergodicity implies that the
distribution of Poincar\'{e} recurrence times is characterized by a
finite first moment, thereby yielding $\mu >2$. According to 
Ref.\cite{allegrini}, $\psi(t)$
is proportional to the second time
derivative of the correlation function $\Phi_{\xi}(t)$ of
Eq.( \ref{asymptotics}). Thus, $\mu = \beta +2$ and the condition $\mu >3$ 
implies $\beta>1$. The values of $\mu>3$, corresponding to a condition where
also the second moment of $\psi(t)$ is finite, means that the ensuing
diffusion process collapses into the attraction basin of the central
limit theorem, and consequently into the dominion of ordinary
statistical mechanics. In conclusion, the stationary condition and our
wish to study anomalous condition force us to consider the interval
defined by Eq.(\ref{anotherimportantcondition}).

The stationary condition involved in the dynamical
theory of diffusion behind Eqs.(\ref{2D(t)})
and (\ref{defineD(t)}) is shared by the mathematical foundation
of the KS entropy, resting on the assumption
that the dynamical system under study
has an invariant measure \cite{pincus}. This
makes it possible to settle the apparent paradox
associated with the use of the map of
Geisel and Thomae \cite{geiselthomae84}
for subdiffusion. It is known \cite{bettin} that
when the map of Geisel and Thomae is adopted
to establish the distribution of sojourn times
on space states rather than on velocity states,
it results in subdiffusion, with the following
time asymptotic expression of the space second moment:
\begin{equation}
 \lim_{t \rightarrow \infty} \langle x^{2}(t)\rangle
 = const \cdot t^{\mu -1},
 \label{bettinsubdiffusion}
 \end{equation}
where $\mu$ is the inverse power index of the function
$\psi(t)$ with
\begin{equation}
1 < \mu < 2.
\label{bettinrange}
\end{equation}
On the other hand, it has been recently
established \cite{condmatt} that according
to the prescriptions of Eq.(\ref{delta(t)})
the map of Geisel and Thomae must be associated
with the entropic index $q$ given by
\begin{equation}
q = 1 + \frac{1}{\mu}.
\label{annamarco}
\end{equation}
This prescription and Eq.(\ref{bettinsubdiffusion})
seem to contradict the simple rule that $q>1$ and
$q<1$ are associated to super and subdiffusion, respectively.
This apparent paradox is settled by noticing that Bettin
et al. \cite{bettin} have demonstrated that the
subdiffusion property of Eq.(\ref{bettinsubdiffusion})
is incompatible with the stationary approach
yielding Eqs.(\ref{2D(t)}) and (\ref{defineD(t)}).

After proving that the stationary assumption is an ingredient
necessary to
satisfactorily relate mechanics to thermodynamics along the lines
established by
Tsallis\cite{CONSTANTINO88,heuristic1,heuristic2,heuristic3},
 we now establish an
important property of subdiffusion: A dynamic
process fulfilling the stationary condition yields
localization within a finite time scale.  We have
seen that, in addition to superdiffusion, another way of breaking the
condition for normal diffusion is given by the property
\begin{equation}
\lim_{t \rightarrow \infty} D(t) = 0.
\label{vanishingdiffusion}
\end {equation}
Note that at $t=0$ the diffusion coefficient $D(t)$  vanishes, due to its
own definition. Thus we expect $D(t)$ to reach the maximum value $D_{max}$
at a finite time
$\tau$, corresponding to the duration of the transition from
 microscopic dynamics to the ``macroscopic'' diffusion regime.
Let us define ($ t > \tau$)
\begin{equation}
R(t) \equiv  \frac{D(t)}{D_{max}}
\label{R(t)}
\end{equation}
and the corresponding localization time:
\begin{equation}
T\equiv \int_{\tau}^{\infty} dt R(t).
\label{localizationtime}
\end{equation}
The case $T =\infty$ corresponds to the asymptotic property of
Eq.(\ref{bettinsubdiffusion}). This is a case of no interest for the
discussion of this paper, focusing on the stationary condition.
Note, however, that the stationary condition implies that use is
made of the correlation
function $\Phi_{\xi}(t)$. The asymptotic behavior of
this correlation function can be derived from Eq.(\ref{defineD(t)}) by
time differentiation. Let us also adopt the following
subdiffusion condition
\begin{equation}
 \lim_{t \rightarrow \infty} \langle x^{2}(t)\rangle
 = const \cdot t^{2 - \beta},
 \label{stationarysubdiffusion}
 \end{equation}
 with
\begin{equation}
1 < \beta < 2.
\label{integrable}
\end{equation}
This is the same subdiffusion regime as that of
Eq.(\ref{bettinsubdiffusion}). The change of notation with the
adoption of the new symbol $\beta$ is due to the fact that
we want now to establish a connection with
a stationary condition and with the correlation
function
$\Phi_{\xi}(t)$. From Eqs. (\ref{2D(t)}) and (\ref{defineD(t)}) we see that
the asymptotic behavior of the correlation function
$\Phi_{\xi}(t)$ is proportional to the second time derivative
of $x^{2}(t)$ of Eq. (\ref{stationarysubdiffusion}). Thus we see that
asymptotic limit of this correlation function
is characterized by negative
tail:
\begin{equation}
\lim_{t\rightarrow\infty}\Phi_{\xi}(t )
= - \frac{const}{t^{ \beta }}.
\label{negativetail}
\end{equation}
This
negative tail would yield a negative contribution to the diffusion
coefficient balancing, in part or totally, the positive contribution
stemming from the short times, thus leading to either normal
diffusion or localization at finite times.
If we want to set the condition of anomalous diffusion
with a vanishing diffusion coefficient,
we are left with a localization process
taking place with the
finite time scale $T$.

In conclusion, the main result of this Section is
that \emph{a dynamic process
with $q<1$ must involve localization with finite time scale}. 
The demonstration proceeds as follows. First we assume
that the property found by the authors of Ref.
Ref.\cite{tsallisbukman96} that $q<1$ implies subdiffusion is of
general validity. The authors base their arguments on
Fokker-Planck-like
equations, and consequently on a probabilistic
view, which is historically dictated by
the need itself of understanding thermodynamics.
We are convinced that a unifying perspective
directly relating thermodynamics to dynamics is possible. On the
basis of this conviction we make the conjecture
that $q<1$ implies subdiffusion even when a totally dynamics
perspective is adopted. We note that the extension of KS entropy
rests on the existence of an invariant distribution and consequently
is compatible with a dynamic approach to diffusion
resting on the stationary assumption. Finally, the adoption of the
dynamic approach to diffusion\cite{mannella} shows that stationary
subdiffusion means localization occurring within a finite time scale.

\section{thermodynamics of fractal dynamics}
The authors of Ref.\cite{jin} expressed the Kolmogorov-Sinai-Tsallis
(KST) entropy in terms of an average over the invariant distribution
$p(x)$ by means of the following expression:
 \begin{equation}
  H_{q}(t) = [1 - k(q) \int dx p(x)^{q} \delta(t,x)^{1-q}]/(q-1).
  \label{fundamentalresult}
  \end{equation}
  Note that the departure point of the theory leading to this
  interesting result is given by a repartition of the phase space of
  the system under study into cells of small size. This is
  mirrored by the fact that $k(q) \equiv  (2/l)^{q-1}$, where
  $2/l$ denotes the size of these cells. The function $\delta(t,x)$
  is defined by Eq.(\ref{delta(t)}) with the dependence on the
  initial condition $x$ made explicit.

 It is worth noticing that the interesting result of
 ref.\cite{jin}, as expressed by
 Eq.(\ref{fundamentalresult}),
 fits the predictions of the more heuristic treatment of
 Refs.\cite{heuristic1,heuristic2,heuristic3}. Let us illustrate that
 this aspect in detail. Let us assume that the function
 $\delta(t)$ of Eq.(\ref{delta(t)}) is expressed by
 \begin{equation}
 \delta(t) = [1 + (1-q) \lambda_{Q} t]^{1/(1-Q)}.
 \label{typicalsensitivity}
 \end{equation}
Note that we are using the capital letter $Q$ rather than
$q$ on purpose. This is because we are assuming that
that only one power law exists, that this is characterized
by the power index $\beta= \frac{1}{(1-Q)}$ and that the resulting $Q$ is
the magic value that the analysis made in terms of the KST ought to
discover. Plugging Eq.(\ref{typicalsensitivity}) into the KST entropy
of Eq.(\ref{fundamentalresult}) we immediately assess that the
KST entropy become a linear function of time only when we assign to the mobile
entropic index $q$ the magic value $q = Q$.

 Note that
 the threshold of chaos is characterized by fractal dynamics. This
 means that system's dynamics significantly depart from the condition
 of full chaos, where\cite{zaslavsky} a nice connection between
 thermodynamics and mechanics can be established using the
 KS entropy. The non-extensive thermodynamics of Tsallis makes it
 possible to extend a thermodynamic
 perspective to the case of fractal dynamics .
We want to point out again that the onset of thermodynamics is
associated, in a full accordance with the traditional
 wisdom\cite{zaslavsky}, to the presence of a time region where
 the generalized form of Kolmogorov entropy grows linearely in time.

 Actually, as earlier pointed out by the authors of
 Refs.\cite{politi,mori}, in the real dynamic cases, the sensivity to
initial condition
 does not result in a
 simple expression as that of Eq. (\ref{typicalsensitivity}). Rather,
  a distribution of power indexes is found. Work in progress of our
  group aims at assessing which value of $Q$ will emerge from the
  search for
  linear increase of $H_{q}(t)$ of Eq.(\ref{fundamentalresult}), if a
  proper average over this distribution of power indices is made.
 Although a final conclusion
 has not yet been reached, we
 are convinced that the prediction  $Q <1$ of
 Refs.\cite{heuristic1,heuristic2,heuristic3} is correct.
  Thus we can restate our convition that the family of logisticlike maps
 studied in Section IV has to produce subdiffusion under the specific
 form of localization with a finite localization time.

\section{The logistic map and the process of dynamical localization
with a finite time scale}
We devote this Section to checking the conjecture that
$Q < 1$ yields a localization process with a finite localization time.
For the reasons explained in Section I, we use
the logistic map  as a diffusion generator. More precisely,
we follow \cite{heuristic3} and we
study
the logisticlike family of maps:
\begin{equation}
y_{n+1}= 1 - \mu | y_{n}|^{z},
\label{logisticmap}
\end{equation}
with
\begin{equation}
z> 1; 0 < \mu \le 2; n = 0,1,2,\ldots; y \in [-1,1].
\label{range}
\end{equation}

First of all, we evaluate numerically the invariant distribution,
a sample of which is illustrated in Fig. 1 for the case $z = 2, \mu =
\mu_{c}(2) = 1.4011551\ldots$. Then we define the fluctuation
\begin{equation}
\xi_{n} \equiv y_{n} - <y_{n}>_{eq},
\label{fluctuation}
\end{equation}
where $<y_{n}>_{eq}$ denotes an average over the invariant distribution.
This makes it possible for us to identify the fluctuation of
Eq.(\ref{fluctuation}) with the diffusion generator of
Eq.(\ref{mainequation}). In fact, it is evident that with
$n\rightarrow \infty$ the discrete variable of Eq.(\ref{fluctuation})
becomes virtually continuous as the variable $\xi$ of
Eq.(\ref{mainequation}).
Then we evaluate $x(t)$ using the discrete counterpart of
Eq.(\ref{x(t)}) with different values of z,
we determine $x(t)^{2}$ and we average it over the set of
initial conditions that in the case $z = 2$ corresponds to the invariant
distribution of
Fig.1.

Actually, we do our calculations for different values of $z$, and for
each $z$ we assign to $\mu$ the corresponding critical value
$\mu_{c}(z)$.
 According to the authors of
 Ref\cite{heuristic3} the best fitting to their numerical result is
 given by:
 \begin{equation}
 Q(z) = 1 - a_{0}/(z-1)^{a_{1}},
 \label{bestfitting}
 \end{equation}
 with $a_{0}= 0.75$ and $a_{1}= 0.60$. This means that with
 increasing $z$ the value of $Q$ tends to the value $Q = 1$,
 signalling the
 condition of standard statistical
 mechanics\cite{list,CONSTANTINO88,heuristic1,heuristic2,heuristic3}.
 This implies that for $z \rightarrow \infty$, the diffusion process
 must tend to become ordinary Brownian diffusion again. Let us see
 how the time evolution of the second moment $M_{2}(t) \equiv
 <x^{2}(t)>$ mirrors this earlier entropic analysis\cite{heuristic3}.
The time evolution of $M_{2}(t)$ is illustrated by Fig.2. We see that
the time evolution of $M_{2}(t)$
shows a scenery distinctly different from that of
ordinary statistical mechanics, corresponding
to a function $M_{2}(t)$  increasing linearly in time. At $z= 2$ the
function $M_{2}(t)$
immediately
settles in a state characterized by wild fluctuations but
 not exceeding a finite upper value. We also see that
 increasing $z$ from  $z =2$ to
 $z = 100$ has the effect of resulting
 in a visible localization onset time. In fact in the case $z=100$
 it takes the system about
 $25$ time steps to settle in another state characterized by wild
 fluctuations around a time indipendent mean value.
 Although the intensity of the fluctuations is larger than in the case
 $z = 2$, the transient nature of this transition to the localized
 state is clear.
 Exploring the time evolution of the logisticlike map of
 Eq.(\ref{logisticmap}) for larger values of $z$ so as to make
 more extended the time necessary to produce localization
 involves some technical problems\cite{capel}. However, on the basis
 of the results illustrated in Fig.2 we make the conjecture that the
 effect of increasing $z$ must be that of making increasingly larger 
 the localization time so that the condition of ordinary statistical
 mechanics $Q(\infty) = 1$ is recovered when the localization time
 becomes infinite.

 We explored another path to full chaos, namely to the entropic
 condition characterized by $Q = 1$. This
 is obtained by keeping $z=2$ and moving $\mu(2)$ from
 the critical value $\mu = \mu_{c}(2) = 1.4011551\ldots$ to the value
 $\mu = 2$, in correspondence of which the logistic map is
 characterized by full chaos\cite{chinesebook}.
 First of all we adopt the same numerical method as that used in
 Refs.\cite{heuristic1,heuristic2,heuristic3} to establish the
 entropic index $Q$ corresponding to a given dynamical condition.
 We use this same method with a changing  value of the control parameter
 $\mu$: we select some
 values of $\mu$ in the interval $[1.4011551\ldots, 2]$ that do not
 correspond to the periodic windows.
 The behavior of $\delta(t)$ in these cases is illustrated by Fig.3 and the 
 second moment of the diffusing distribution is illustrated by Fig.4.

 These numerical results are somewhat unexpected. In fact, we would
 have
 expected the localization time to become increasingly larger till to
 diverge at the condition of full chaos. On the contrary, we see that there
 exists a localization state regime. This is of infinite time
 duration at the critical threshold $\mu = \mu_{c}$. As the control
parameter $\mu$
 exceeds this critical value the localization state become unstable.
The closer and the closer the control parameter $\mu$ to the full chaos
condition, the shorter and the shorter the lifetime of this localized state.
 In the literature of localization processes it is possible to find a
 discussion of the effect produced by the environmental fluctuations
 on the phenomenon of localization exhibited by the kicked rotator
 (see, for instance, Ref.\cite{graham}). This is a kind of
 localization breakdown taking place as soon as the interaction with
 environment is switched on. The diffusion coefficient of the
 ensuing process of ordinary diffusion is proportional to
 the square of the intensity of the coupling between system and environment.
 Here, on the contrary, we see that the localization breakdown occurs
 in time, at times of increasingly larger duration as the control parameter
 approaches the threshold value $\mu_{c}$.

In conclusion, it seems that the effect revealed by the numerical
calculations of this paper is related to the conjecture of Ref.\cite{jin}
 that the entropic index Q is time dependent. As a result of a proper average,
made on the
the crowding index distribution of Ref.\cite{beck} rather than on the 
coordinate $x$, we expect that Eq.(\ref{fundamentalresult}) results in a 
linear in time entropy increase at
a "magic" entropic value Q. This value, in turn, is not the same at all
times. This is equivalent to saying that, in a sort of effective sense, the
form of Eq.(\ref{typicalsensitivity}) is replaced by

 \begin{equation}
 \delta(t) = [1 + (1-Q(t)) \lambda_{Q(t)} t]^{1/(1-Q(t))}.
 \label{agingeffect}
 \end{equation}

In accordance with the main result of this paper, let us assume that $ Q(t) <
1$ means localization occurring within a finite time scale, and, as shown by
Fig.2, almost instantaneously at $z=2$.  Then the results of Fig.4 can be
immediately accounted for if we assign to $Q(t)$ the following time behavior:
$Q(t)$ keeps a value smaller than the ordinary prescription $Q = 1$ for an
extended time interval corresponding to the time duration of the localized
states of Fig.4, then $Q(t)$ makes an abrupt jump to the traditional value $Q
= 1$.
Note that Fig.3, and other figures as well, concerning different
values of system's parameters and not shown here, exhibit an interesting
property worth of a comment. The overall time behavior of $\delta(t)$ looks
like exponential in accordance with the fact that the localization
breakdown implies that the standard form of sensitivity to the initial
conditions is recovered. However, a more careful analysis of the numerical
results shows that the exponential increase of $\delta(t)$ and the
multifractal oscillations of Refs. \cite{heuristic1,heuristic2,heuristic3}
 coexist in the same curve. The
multifractal oscillations last for a finite amount of time at a given
level, with a behavior very similar to that illustrated in Refs. 
\cite{heuristic1,heuristic2,heuristic3}. At
a given time a sort of abrupt transition to another level, with the same
kind of multifractal oscillations, take place. The overall behavior looks
like that exponentiallike behavior of Fig.3. We think that the abrupt
transition from a localized state condition to the regime of Brownian
diffusion, shown in Fig.4, is  a consequence of the coexistence of these
two properties. Thus we make the following attractive conjecture: The
occurrence of a localization process with a finite time scale, as a result
of the stationary condition and of an entropic index $Q <1$, is closely
connected to the multifractal nature of the oscillations exhibited by
$\delta(t)$, as revealed by the analysis of Ref. 
\cite{heuristic1,heuristic2,heuristic3}. 
In these papers no significant attention was devoted to the 
physical meaning of these
oscillations. We hope that the numerical results of this Section might
trigger further research work aiming at establishing the nature of this
attractive connection.

\section{Concluding remarks}
The so called KST entropy shares with the conventional KS entropy the
stationary condition, which is made evident by
Eq.(\ref{fundamentalresult}), resting in fact on an average over the
invariant distribution. The findings of Ref.\cite{tsallisbukman96}
refer to a probabilistic level of description, preceding the first
attempts at establishing a dynamic foundation of non-extensive
thermodynamics\cite{heuristic1,heuristic2,heuristic3,condmatt}.
We make the plausible assumption that the unification of dynamics and
thermodynamics\cite{polish,bianucci} can be extended. As ambitious as
this task is, this assumption is plausible, and the first results
obtained\cite{condmatt} are very encouraging. Consequently, it is
reasonable to make the conjecture that the results of
Ref.\cite{tsallisbukman96} have a general validity and apply also to
merely dynamic processes, described with no use of probabilistic
ingredients. This means that a dynamic process characterized by $Q <1$
is expected to yield subdiffusion.
On the other hand, according to the compelling results of Section III a
process of subdiffusion generated by stationary fluctuations must
lead to localization within a finite time scale. All these arguments
lead us to make the plausible conjecture that dynamics with $Q<1$ generate
localization, and that the localization time is finite.

This is not a rigorous demonstration. However, the numerical results
illustrated in Section IV fully support this conjecture and, in our
opinion, should serve the useful purpose of stimulating a search for a
more rigorous demonstration.

\begin{figure} [h]
\centerline{\epsfig{figure=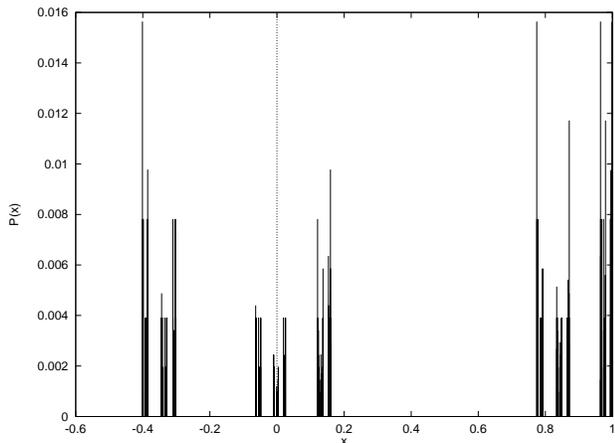,width=3.3 in}}
\caption{Fig1.  The invariant distribution of the logistic map of
Eq.(\ref{logisticmap}), with $z = 2$, at the critical value $\mu =
1.4011551\ldots$. }
\label{fig1}
\end{figure}

\begin{figure}[h]
\centerline{\epsfig{figure=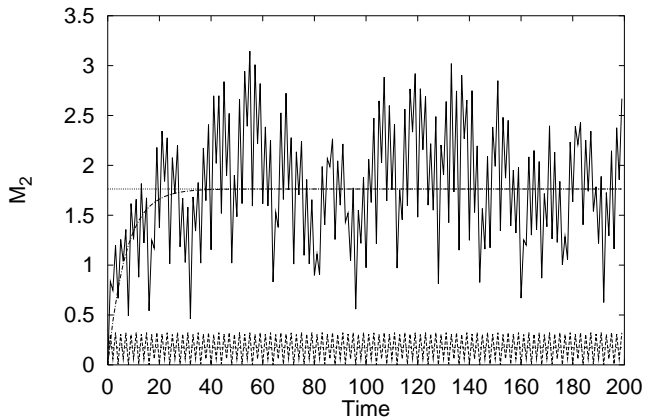,width=3.3 in}} 
\caption{Fig2. The second moment $M_{2} \equiv <x^2>$ of the logistic
map of Eq.(\ref{logisticmap}) as a function of time. We show the case $z = 2$ 
(bottom curve) and the case $z = 100$ (uppercurve). The bottom dashed line and
 the upper dottel line are eye guides illustrating that in both cases the 
second moment fluctuations take place around time independent mean values. The
 dot-dash line is an eye guide illustrating the existence of a finite time of
transition to the localized state in the case $z = 100$.}
\label{fig2}
\end{figure}

\begin{figure}[h] 
\centerline{\epsfig{figure=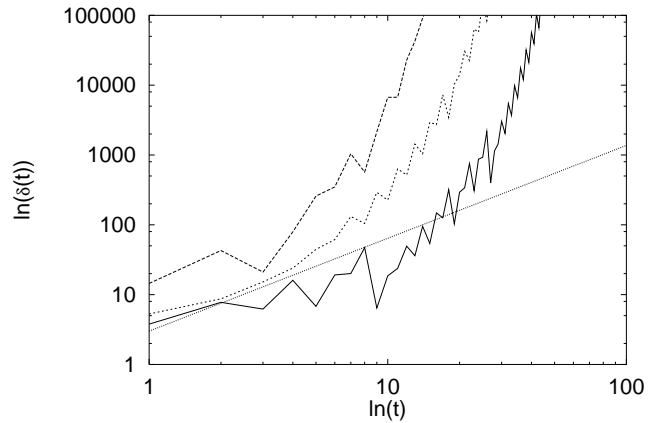,width=3.3 in}}
\caption{Fig3. The function $\delta(t)$ of Eq.(\ref{delta(t)}) of the 
logistic map of Eq.(\ref{logisticmap}) as a function of time. We show the 
case $z = 2$ with $\mu =1.45$ (bottom), $\mu =1.55$ (middle) and the case 
$\mu = 1.95$ (upper curve).}
\label{fig3}
\end{figure}
\begin{figure}[t] 
\centerline{\epsfig{figure=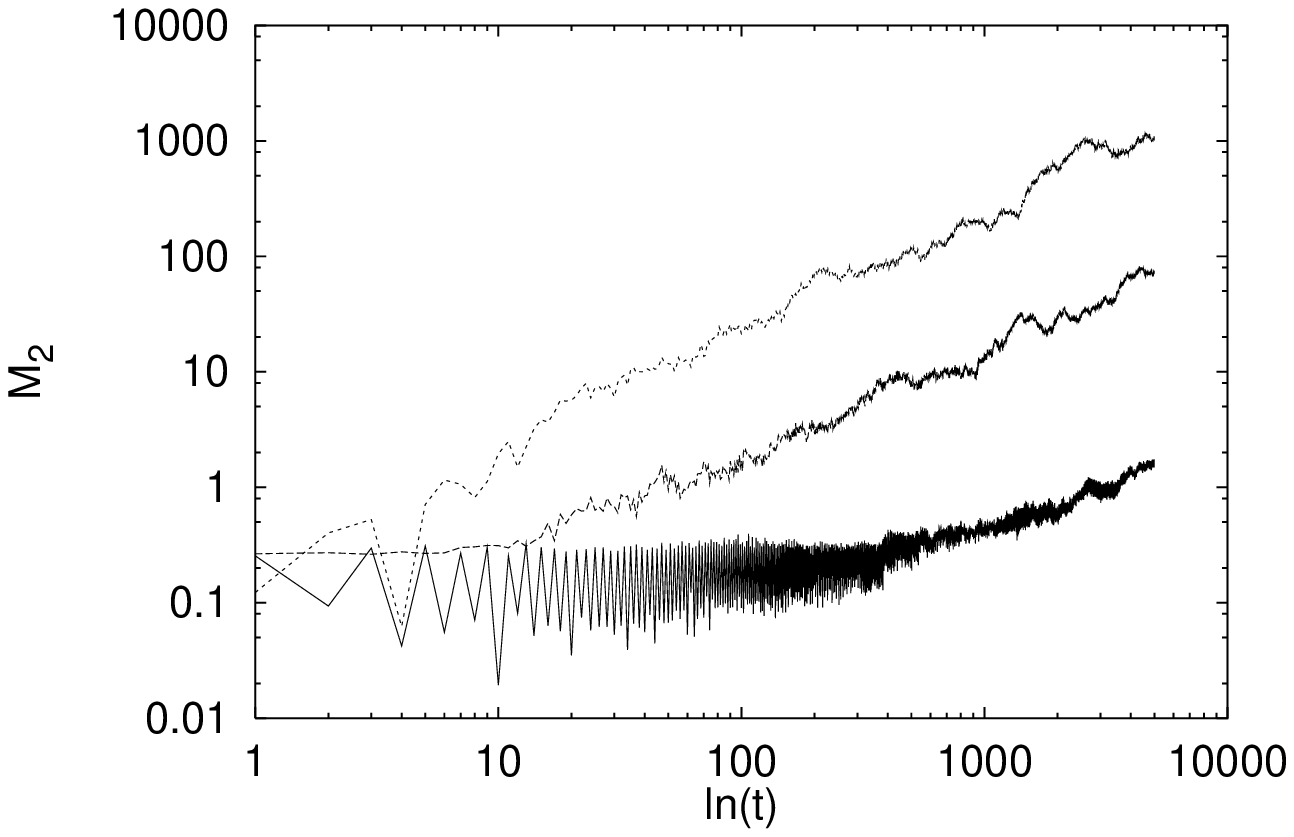,width=3.3 in}}
\caption{Fig4.The second moment $<x^2>$ of the logistic map of
Eq.(\ref{logisticmap}) as a function of time. We show the case $z = 2$ with
$\mu = 1.45$ (bottom), $\mu =1.55$ (middle)  and the case $\mu=1.95$ 
(upper curve).}
\label{fig4}
\end{figure}

\end{document}